\documentclass [structabstract]{aa}
\usepackage{txfonts}
\usepackage{natbib}
\usepackage{graphicx}
\usepackage{color}
\usepackage{rotating}

\bibpunct{(}{)}{;}{a}{}{,}

\begin{document}

\title{VLT/NACO astrometry of the HR\,8799 planetary system
  \thanks{Based on observations collected at the European Southern Observatory, Chile, under programme ID 084.C-0072}}
\subtitle{L$^{\prime}$-band observations of the three outer planets}

\author{C. Bergfors\inst{1}\thanks{Member of the International Max Planck Research School for Astronomy and Cosmic Physics at the University of Heidelberg}
  \and W. Brandner\inst{1}
  \and M. Janson\inst{2}
   \and R. K\"ohler\inst{1,3}
  \and T. Henning\inst{1} 
}	
\institute{Max-Planck-Institut f\"ur Astronomie, K\"onigstuhl 17, 69117
  Heidelberg, Germany \\
  \email{bergfors@mpia.de}
  \and University of Toronto, Dept. of Astronomy, 50 St George Street, Toronto, ON, M5S 3H8, Canada
  \and Landessternwarte, Zentrum f\"ur Astronomie der Universit\"at Heidelberg, K\"onigstuhl, 69117 Heidelberg, Germany
}

\date{Recieved 11 January 2011/
			Accepted 16 February 2011}

\abstract{HR\,8799 is so far the only directly imaged multiple exoplanet system. The orbital configuration would, if better known, provide valuable
insight into the formation and dynamical evolution of wide-orbit planetary systems.}
{We present data which add to the astrometric monitoring of the planets HR\,8799\,b, c and d. We investigate how well the two simple cases of (i) a
circular orbit and (ii) a face-on orbit fit the astrometric data for HR\,8799\,d over a total time baseline of $\sim2$\,years. 
}
{The HR\,8799 planetary system was observed in L$^\prime$-band with NACO at VLT.}
{The results indicate that the orbit of HR\,8799\,d is inclined with respect to our line of sight, and suggest that the orbit is slightly eccentric
or non-coplanar with the outer planets and debris disk.}
{}


\maketitle

\titlerunning{VLT/NACO astrometry of the HR\,8799 system}
\authorrunning{C. Bergfors et al.}

\section{Introduction}
As the first and so far only directly imaged multiple exoplanet system, the HR\,8799 system carries the promise of providing valuable insight into the 
structure and 
characteristics of planetary systems. While more than 500 extrasolar planets have now been discovered, most have been found by radial velocity and 
transit searches; the sample of known exoplanets is thus heavily biased towards short-period planets. 
Directly imaged giant extrasolar planets provide a necessary complement to these indirect detection techniques for a full picture of the 
characteristics of planets, and are crucial for theories of planet formation.

Challenging as it may be to directly image planets, whose relatively faint light is easily lost in the bright stellar glare, several 
confirmed companions are known. As of November 2010, 7 planetary mass objects belonging to stars, including the quadruple-planet HR\,8799 system, 
have been discovered with direct imaging (Fomalhaut b, \citet{Kalas2008}; $\beta$ Pic b, 
\citet{Lagrange2009, Lagrange2010}; 1RXS J160929.1-210524 b, \citet{Lafreniere2008, Lafreniere2010}, and HR\,8799 bcde, \citet{Marois2008,Marois2010}). 
The HR\,8799 system is 
especially interesting since its multiple planet configuration allows for comparison of the characteristics of planets within the same environment of 
formation and evolution. 
The star is a young \citep[30-160\,Myr, ][]{Marois2008} A5\,V star at at distance of 39.4\,pc from the Sun \citep{vanLeeuwen2007},
surrounded by a debris disk \citep{Rhee2007, Su2009}. Three of the planets in the system, HR\,8799\,b, c and d, were discovered in 2008 and an additional 
planet, 
HR\,8799\,e, in 2010 \citep{Marois2008,Marois2010}, adding up to at least four giant planets of masses 7-10\,$\rm M_J$ at projected separations 14.5, 24, 38 and
68\,AU from the central star.

The astrometric analysis of the three outermost planets at the time of their discovery provided evidence that the planets are co-moving with the star, 
and suggested that their orbits are almost circular and seen close to face-on.
However, dynamical modelling of the HR\,8799 system has since shown that this initially presumed configuration of orbits is unlikely for reasons of 
orbital stability of the system \citep{Fabrycky2010, Reidemeister2009, Gozdziewski2009}.  \citet{Fabrycky2010} found that for the masses 
derived by \citet{Marois2008} and circular, face-on orbits, the system would become unstable at an age of only $\sim10^5$ years, i.e. significantly 
younger than its assumed present age. 
Stable mean motion resonance configurations
were found for the three outer planets known at the time by \citet{Fabrycky2010, Gozdziewski2009} and \citet{Reidemeister2009}, and 
\citet{Marois2010} found stable resonant
configurations including also the fourth planet.

In this paper, we present astrometric measurements of the three outermost planets in the HR\,8799 system. The observations were obtained with NACO at 
VLT in September 2009
 -- one year after the discovery of planets b, c and d. 
We investigate how well two simple models with (i) a circular orbit and 
(ii) a face-on orbit fit the astrometric data when the new observations are included.

\section{Observations and data reduction}

Acquisition images of HR\,8799 in L$^\prime$-band were obtained with NACO/VLT \citep{Lenzen2003, Rousset2003}
on the nights of October 5 and 6, 2009, as part of the observation programme 084.C-0072, in which a spectrum of the 
planet HR\,8799\,c was obtained \citep{Janson2010}. The observations were acquired in cube-mode and consisted on each night of 2 sets of 10 data cubes, 
one set taken at the default orientation with north up and one rotated by 
33\degr. The frames were obtained with the purpose of checking the alignment for slit orientation. 
Each cube
contained 749 usable frames
on October 5 and 1499 frames on October 6 with individual integration 
time 20.2\,ms,
yielding a total integration time of 15\,s and 30\,s respectively per data cube. 

Images of the astrometric binary HD 211742 obtained in September 2009 were retrieved from the ESO (VLT) archive and used for calibration of plate scale 
and true north orientation
of the detector. We derive a field rotation of -0.6$\degr\pm0.2\degr$ and plate scale of 27.1\,mas/px, consistent with the 27.2 mas/px of the L27 camera
described in the NACO Usermanual, 
assuming a systematical error of $0.2\degr$ and 0.3 mas/px \citep[$1\%$ of the pixel scale, see ][]{Koehler2008}.

\begin{figure}
 \resizebox{\hsize}{!}{\includegraphics{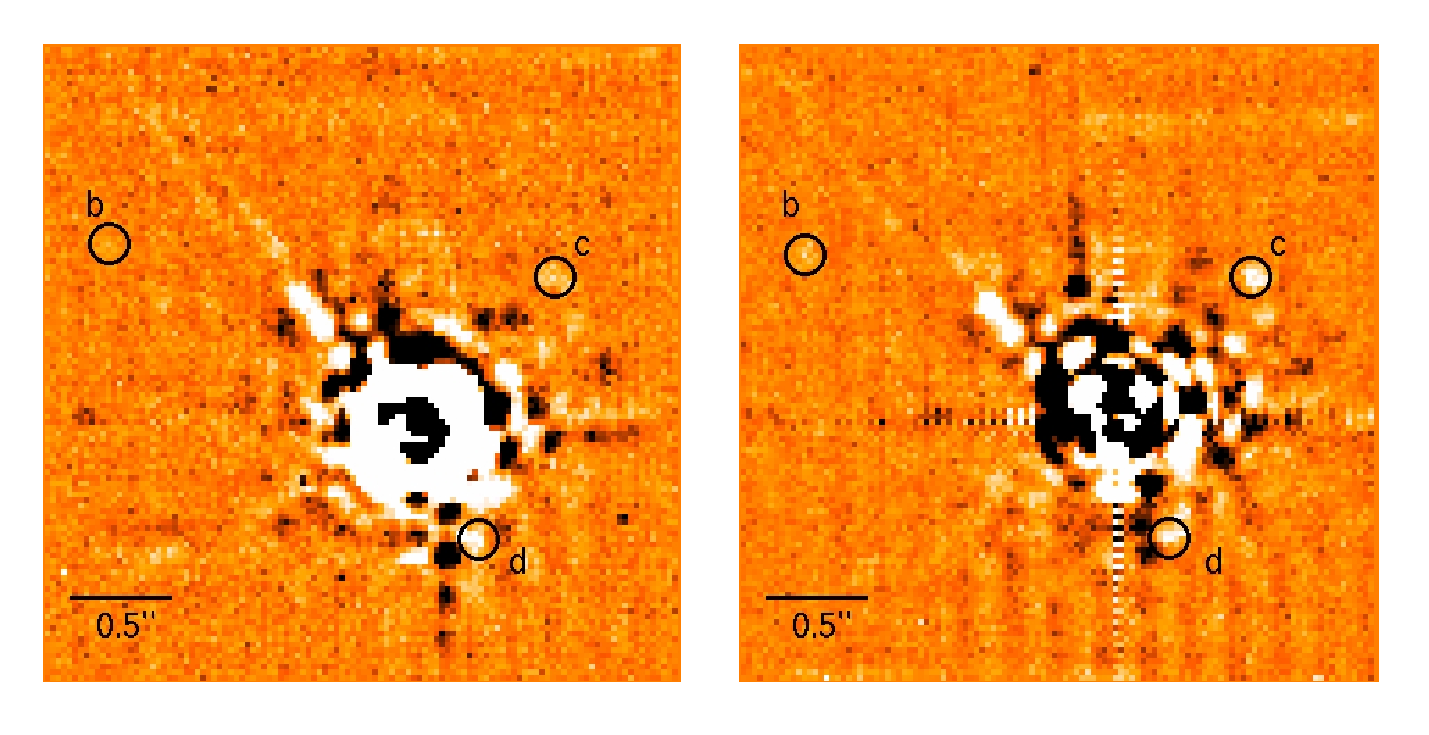}}
    \caption{NACO L$^{\prime}$-band images of the HR\,8799 system. Rotated and subtracted 3$^{\prime\prime}$ x 3$^{\prime\prime}$ images from 
observations on October 5 (left) and October 6 (right)
2009. North is up and east is to the left. Scaling is linear. }
 \end{figure}

\begin{figure}
  \resizebox{\hsize}{!}{\includegraphics{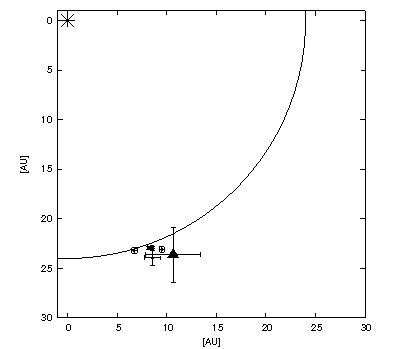}}
    \caption{Observations of HR\,8799\,d (see Table 1). A distance to the star of 39.4\,pc is assumed, and the nominal circular, face-on orbit at 24 AU 
is overplotted. The star is marked with an asterisk at (0,0). The triangle marks our NACO L$^{\prime}$ observation.}
 \end{figure}

The data reduction was performed using IRAF and IDL. Skyframes were constructed by averaging the sum frame of 10 data cubes obtained at the 
approximate same time 
and airmass with the star in different
dither positions, rejecting the 2-3 highest values at each pixel in order to remove the stellar flux. The skyframes were subtracted from each 
frame in the cubes, and bad pixels were replaced by the mean value of neighbouring pixels.
The frames were aligned by fitting a 2-D Gaussian to the star and measuring the centroid position, and then co-added to remove residual tip-tilt
between individual frames of each data cube and produce one image per rotation angle for each night. Frames of poor quality 
or with the target too close to the detector edge, hence cutting out the planets, were rejected, resulting in a combination of 4-7 data cubes for each
final image (4 images in total, one for each rotation angle on each night). Unsharp masking was used on the co-added frames by smoothing one
version of the image using a boxcar average with 15 pixels and subtracting the smoothed image from the original.
The positions of the planets and central star were determined using the IRAF
\textit{imexamine} task. While the brightest planet, HR\,8799\,c, was clearly detectable in all four images, the position of the fainter b-planet could
only be determined from 3 measurements. Speckle contamination obscured planet d in the $33\degr$ rotated images and
the position could thus only be determined from 2 frames. 
The reported position of planet e coincides with the third diffraction ring and is not detected with significant counts. 
Fig. 1 shows rotated and subtracted images from 
both nights,

When imaging sources with vastly different spectral energy distributions through a broadband filter at an airmass $>1.0$, the effect of differential
atmospheric refraction on the relative astrometry has to be considered \citep[see, e.g., ][]{Helminiak2009}. An effective wavelength $\lambda_{\rm eff}=3.777\mu$m for our NACO L$^\prime$
imaging observation of the star HR\,8799 was computed by convolving a spectrum of a star of similar spectral type from the IRTF SpeX spectral
library \citep{Cushing2005, Rayner2009} with the L$^\prime$ transmission curve from the NACO Usermanual. For the three exoplanets, an effective
wavelength of $3.905\pm0.010\,\mu$m was computed using model spectra \citep[][ and priv. comm.]{Burrows2006} for an effective
temperature of 1100\,K and log g in the range 3.0 to 5.0.

As a consistency check, we also computed the effective wavelength for L$^\prime$ imaging observations of Jupiter based on the IRTF SpeX library, which
resulted in $\lambda_{\rm eff}=3.907\,\mu$m. Hence, once effective temperatures in the atmospheres of substellar objects are low enough to allow for
the pronounced presence of water, methane and CO molecular absorption bands, the $\lambda_{\rm eff}$ values for L$^\prime$ observations seem to show
little dependence on effective temperature and surface gravity.

Next, we used the model fits for the refractive index of humid air in the infrared as computed by \citet{Mathar2007} to estimate the amplitude of
differential atmospheric refraction of HR\,8799 and its exoplanets. We found that this effect caused a shift of $\approx5-6$\,mas of the planet
positions along the parallactic angle. The positions of the planets relative to the star are presented in Table 1, 
together with all previously published position measurements. 
The errorbars were estimated by introducing artificial structures in the form of Gaussians with the same approximate peak flux and the same 
separations from the star as the real
planets but at different angles. 
The deviations from the known positions were measured for a set of 
3 different angles per planet and image. The standard deviations of all measurements for each ``fake planet'' are the estimated errors, mainly
due to residual speckles for the two innermost planets.

\section{Results and discussion}
\subsection{Astrometric measurements of HR\,8799\,b, c and d}
The projected orbital motions of the two outermost planets b and c are slow \citep[the periods are $P_b\sim460$ and $P_c\sim190$ years respectively, ]
[]{Marois2008},
and the nominal mostly circular, face-on orbits are still consistent with observations when our data are added to the previously published 
astrometry. However, observations in 2008 and 2009 of planet d by \citet{Currie2011} and \citet{Hinz2010}\footnote[1]{The MMT/Clio data by 
\citet{Hinz2010} were rereduced by \citet{Currie2011}. We have adopted the \citet{Currie2011} astrometry for this analysis.} together with our 
observations suggest that the orbit of this planet might be eccentric and/or inclined.
All published astrometric measurements of HR\,8799\,d are plotted in Fig. 2.

\subsection{Testing the cases of $i=0$ and $e=0$ for HR\,8799\,d}
We now want to test if the observed changes in position angle and separation are in agreement with different models. \citet{Fabrycky2010} investigated
the stability of different possible orbital families for the HR\,8799 system. Many of these are for circular ($e=0$) or face-on ($i=0$) 
orbits.
We modelled these two simple cases for 
HR\,8799\,d from the astrometric observations over a $\sim2$\,year time baseline. 
The nominal system mass of 1.5\,M$_{\sun}$  \citep{Marois2008} was used to compute
orbital period from semi-major axis.
We did not attempt to fit  
model orbits that are both inclined and eccentric, since this would  
add two free parameters and reduce the statistical significance of the result.
For the following orbital fits we assumed errors as given in the literature and listed in Table 1.
We note, however, that while the literature measurements are probably subject to systematic errors of the same order as the ones present in the
VLT/NACO set, systematic errors (plate scale, detector orientation) between the different telescopes and instruments used are not 
necessarily included in these error bars.

\begin{figure}
  \resizebox{\hsize}{!}{\includegraphics{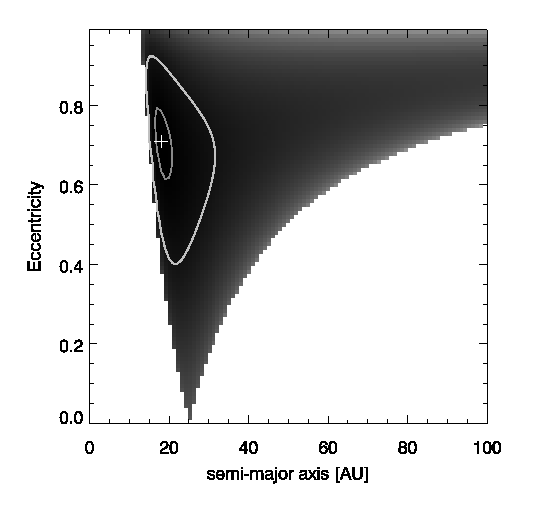}}
    \caption{$\chi^2$-fit of eccentricity and semi-major axis for the orbit of HR\,8799\,d assuming $i=0$ (face-on orbit). 
The best fit is marked by a cross and the 
  contour lines show the $68.3\%$ and $99.73\%$ confidence regions. 
}
 \end{figure}

\begin{figure}
  \resizebox{\hsize}{!}{\includegraphics{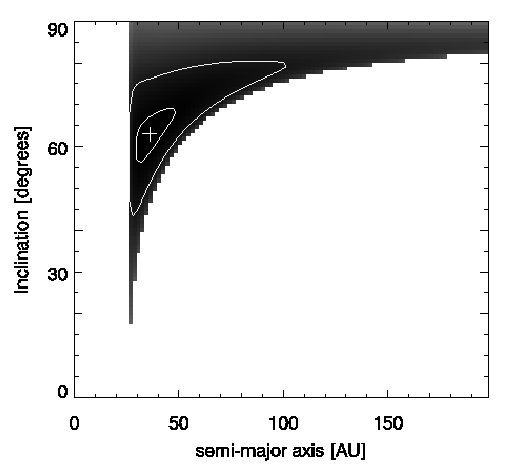}}
    \caption{$\chi^2$-fit of inclination and semi-major axis for the orbit of HR\,8799\,d, assuming $e=0$ (circular orbit).
The best fit is marked by a cross, and the 
  contour lines show the $68.3\%$ and $99.73\%$ confidence regions.}
 \end{figure}

\subsubsection{Face-on orbit model for HR\,8799\,d}
Assuming a face-on orbit ($i=0$) for HR\,8799\,d, 
the $\chi^2$ was computed for a grid of 100 semi-major axes 
and 100 eccentricities and is shown in Fig. 3. The cross marks the best fit
and the contour lines surround the $68.3\%$ and $99.73\%$ confidence regions. We find that a face-on orbit 
must have eccentricity $e\ga0.4$ to fit the observations with $99.73\%$ confidence.
The system should have become dynamically stable at the assumed age of 60\,Myr, and
one of the stability criteria of \citet{Fabrycky2010} is that the orbits of the planets do not cross: $a_d(1+e_d)<0.85a_c(1-e_c)$. 
Our derived $99.73\%$ confidence minimum eccentricity of $e\approx0.4$
corresponds to an apastron distance of $r_{ap}(d)\approx34$\,AU
for the nominal semi-major axis of planet d at $a_d=24$\,AU, 
thereby violating the mentioned stability criterium for the nominal orbit of planet c ($a_c=38$\,AU).
The stability criterium cannot be fulfilled with respect to both the outer planet c and the inner planet e even if the orbits of both c and e are
perfectly circular.
A face-on orbit is, with the astrometric data points taken at face value, unstable because of the high eccentricity, and not likely to represent 
the true orbit of the planet.
\begin{table*}
\caption{Relative positions of the HR\,8799 planets.}
\label{table:1}
\centering
\begin{tabular}{l l l l l r}
\hline\hline
Epoch & HR 8799b & HR 8799c & HR8799d & HR8799e & Ref.\\
 & $\Delta\alpha$, $\Delta\delta$ (arcsec) & $\Delta\alpha$, $\Delta\delta$ (arcsec) & $\Delta\alpha$, $\Delta\delta$ (arcsec) & $\Delta\alpha$, $\Delta\delta$ (arcsec) &\\
\hline
1998.83	&	1.411$\pm0.009$, 0.986$\pm0.009$ 	&	...					&	...					&	...					&	1	\\
2002.54	&	1.481$\pm0.023$, 0.919$\pm0.017$	&	...					&	...					&	...					&	2	\\
2004.53	&	1.471$\pm0.005$, 0.884$\pm0.005$	&	-0.739$\pm0.005$, 0.612$\pm0.005$ 	&	...					&	...					&	3	\\
2007.58	&	1.522$\pm0.003$, 0.815$\pm0.003$	&	-0.672$\pm0.005$, 0.674$\pm0.005$	&	-0.170$\pm0.008$, -0.589$\pm0.008$	&	...					&	4	\\
2007.81	&	1.512$\pm0.005$, 0.805$\pm0.005$	&	-0.674$\pm0.005$, 0.681$\pm0.005$	&	...					&	...					&	3	\\
2008.52	&	1.527$\pm0.004$, 0.799$\pm0.004$	&	-0.658$\pm0.004$, 0.701$\pm0.004$	&	-0.208$\pm0.004$, -0.582$\pm0.004$	&	...					&	3	\\
2008.61	&	1.527$\pm0.002$, 0.801$\pm0.002$	&	-0.657$\pm0.002$, 0.706$\pm0.002$	&	-0.216$\pm0.002$, -0.582$\pm0.002$	&	...					&	3	\\
2008.71	&	1.528$\pm0.003$, 0.798$\pm0.003$	&	-0.657$\pm0.003$, 0.706$\pm0.003$	&	-0.216$\pm0.003$, -0.582$\pm0.003$	&	...					&	3	\\
2008.89	&	1.532$\pm0.02$, 0.796$\pm0.02$		&	-0.654$\pm0.02$, 0.700$\pm0.02$ 	&	-0.217$\pm0.02$, -0.608$\pm0.02$	&	...					&	5	\\
2009.58	&	...					&	...					&	...					&	-0.299$\pm0.019$, -0.217$\pm0.019$	&	6	\\
2009.58	&	...					&	...					&	...					&	-0.303$\pm0.013$, -0.209$\pm0.013$	&	6	\\
2009.62	&	1.536$\pm0.01$, 0.785$\pm0.01$		&	...					&	...					&	...					&	5	\\
2009.70	&	1.538$\pm0.03$, 0.777$\pm0.03$		&	-0.634$\pm0.03$, 0.697$\pm0.03$		&	...					&	...					&	5	\\
2009.76	&	1.535$\pm0.02$,0.816$\pm0.02$		&	-0.636$\pm0.04$, 0.692$\pm0.04$		&	-0.270$\pm0.07$, -0.600$\pm0.07$	&	...					&	7	\\
2009.77	&	1.532$\pm0.007$, 0.783$\pm0.007$ 	&	-0.627$\pm0.007$, 0.716$\pm0.007$	&	-0.241$\pm0.007$, -0.586$\pm0.007$	&	-0.306$\pm0.007$, -0.217$\pm0.007$	&	5	\\
2009.83	&	...					&	...					&	...					&	-0.304$\pm0.010$, -0.196$\pm0.010$	&	6	\\
2010.53	&	...					&	...					&	...					&	-0.325$\pm0.008$, -0.173$\pm0.008$	&	6	\\
2010.55	&	...					&	...					&	...					&	-0.324$\pm0.011$, -0.175$\pm0.011$	&	6	\\
2010.83	&	...					&	...					&	...					&	-0.334$\pm0.010$, -0.162$\pm0.010$	&	6	\\

\hline
\end{tabular}
\\
\tablebib{(1)~\citet{Lafreniere2009}; (2) \citet{Fukagawa2009}; (3)  \citet{Marois2008}; (4) \citet{Metchev2009};
(5) \citet{Currie2011}; (6) \citet{Marois2010}; (7) This work. Only statistical errors have been considered in the table.}
\end{table*}

\subsubsection{Circular orbit model for HR\,8799\,d}
The case of zero eccentricity ($e=0$) was considered by varying the orbital inclination and semimajor axis in the same way as described above. 
Figure 4 shows the $\chi^2$ as a function of inclination and semi-major axis for a circular orbit.
We find that the inclination is greater than $43\degr$ within
the $99.73\%$ confidence limits, with a best fit of $i=63\degr$ and $a=36$\,AU.
This is consistent with the asteroseismic constraints on the stellar rotational inclination of $i\ga40\degr$, with a best fit of
$i=65\degr$ derived by \citet{Wright2011}. However,
this inclination is higher than what has been derived for the orbit of planet b from a 10 year baseline of observations 
\citep[$i\sim13-23\degr$ for a circular orbit, ][]{Lafreniere2009}, and also for the debris disk 
\citep[$3\sigma$ upper constraint, $i_{disk}<40\degr$, ][]{Moro-Martin2010}. 
The observed configuration with four 
visible planets around the star at very different position angles supports a low inclination, if the orbits are coplanar. 
With the astrometric data points taken at face value, the high inclination from the fit with $e=0$ thus 
indicates that the orbital eccentricity of HR\,8799\,d is non-zero, or that the planetary orbits are non-coplanar.

\section{Conclusions}
The initial astrometric analysis performed by \citet{Marois2008} suggested mostly face-on and circular orbits for the three planets of HR\,8799
known at the time.
The orbital periods of the outermost planets b and c are of the order of hundreds of years, and with our additional astrometric measurements 
of these planets the orbits are still consistent with the nominal orbits.

Purely circular, face-on and coplanar orbits have been shown 
to be an unlikely configuration for reasons of dynamical stability \citep{Fabrycky2010, Gozdziewski2009, Reidemeister2009}.
Our analysis of the orbit of HR\,8799\,d implies that such a configuration is 
also inconsistent with the astrometric observations when recent observations by \citet{Hinz2010}, \citet{Currie2011} and our NACO data are included.
For a purely face-on orbit, the eccentricity of HR\,8799 d is $e\ga0.4$ within 99.7\% confidence. The system is not stable for such high $e$ and the 
orbit is hence likely to be inclined with respect to our line of sight. In the case of a purely circular orbit we find that the inclination 
is $i>43\degr$ within the 99.7\% confidence limits. While the astrometric data is still limited, this result agrees well with recent constraints on the
the stellar rotational inclination \citep{Wright2011}, but not with other measurements of the orbital inclination of planet b and the debris disk.
The current astrometric data still allow for different orbital planes for the individual planets and the debris disk (and hence for non-coplanar orbits).
Continued astrometric monitoring over a longer time baseline is required in order to put stronger constraints on the orbits of the HR\,8799 planets.

\begin{acknowledgements}
We would like to thank Adam Burrows for providing model spectra,
which enabled us to quantify the effect of differential atmospheric refraction.
\end{acknowledgements}

\bibliographystyle{aa}
\bibliography{Bergfors_references.bib}

\end{document}